# First-Principles Study of Fe Adsorption and Its Effects on the Mechanical and Electrical Properties of Monolayer and Bilayer Biphenylene Networks

Xiao-Ke Zhang, Zheng-Zhe Lin*

*School of Physics, Xidian University, Xi'an 710071, China*

* Corresponding Author. E-mail address: zzlin@xidian.edu.cn

**Abstract** – Biphenylene network (BPN) is a two-dimensional carbon allotrope that exhibits promising potential for applications across a wide range of fields. In this work, we systematically investigated the adsorption characteristics of Fe atoms on monolayer and bilayer BPN using first-principles calculations. Structural optimization and adsorption energy analysis reveal that the average Fe adsorption energy on monolayer BPN generally becomes more negative with increasing Fe loading, although nonmonotonic variations occur at higher coverages. The optimized structures reveal increasingly pronounced Fe aggregation and structural rearrangements at higher coverages, indicating a transition toward three-dimensional Fe growth on the BPN surface. For bilayer BPN, the energetically preferred adsorption site for Fe atom is located at the center of the interlayer four-membered ring, with an average adsorption energy of −4.3 eV for a single Fe atom. Mechanical properties are further evaluated for pristine and Fe-decorated BPN. The results demonstrate that monolayer and bilayer BPN possess relatively high in-plane Young's and shear moduli, indicative of excellent in-plane mechanical stability. For monolayer BPN, Fe adsorption induces nonmonotonic variations in the in-plane Young's moduli, with no systematic enhancement in stiffness as the Fe coverage increases. At several Fe coverages, the in-plane Young's moduli are reduced relative to pristine BPN. In contrast, the out-of-plane mechanical response of bilayer BPN is significantly affected by Fe incorporation. The effective out-of-plane elastic constant $C_{33}$ of pristine bilayer BPN is calculated to be 24.59 GPa, indicating relatively weak interlayer interactions and facile deformation along the out-of-plane direction. Notably, this property can be substantially enhanced by interlayer Fe adsorption, with $C_{33}$ increasing dramatically to 515.63 GPa upon an Fe/C ratio of 25 %. The transport calculations reveal pronounced anisotropy in the conductivity normalized by the relaxation time $\sigma / \tau$, with markedly different values along the two in-plane crystallographic directions.

## 1. Introduction

The biphenylene network (BPN) is a recently emerged two-dimensional carbon allotrope that has attracted considerable attention in recent years. Unlike graphene, which is composed exclusively of hexagonal rings [1-3], BPN consists of a periodic arrangement of four-, six-, and eight-membered carbon rings [4]. This unconventional multi-ring topology endows BPN with distinct geometric configurations, non-uniform local strain distributions, and unique electronic structures that differ fundamentally from those of conventional two-dimensional carbon materials. As a result, BPN has demonstrated promising physicochemical properties with potential applications in diverse fields, including heterogeneous catalysis and as anode materials for rechargeable batteries [5-9]. Compared to two-dimensional carbon materials composed of a single type of ring, the diverse ring motifs in BPN provide a wider variety of adsorption sites, thereby offering greater flexibility for tuning its structural and functional properties via the incorporation of foreign atoms. In addition, carbon atoms in BPN adopt an $sp^2$ hybridization configuration, where each carbon atom forms σ bonds with three neighboring carbon atoms. The strong in-plane covalent bonding within the carbon framework suggests that BPN possesses excellent mechanical strength and chemical stability [10].

The adsorption and doping of transition metal atoms on two-dimensional materials represent effective strategies for tailoring their electronic structure, magnetic properties, catalytic activity, and interfacial stability. Among various transition metals, Fe atoms are particularly attractive due to their partially filled 3d orbitals, high chemical reactivity, and intrinsic magnetic characteristics, making them a representative system for investigating adsorption behavior and property modulation mechanisms on low-dimensional surfaces [11-14]. Furthermore, Fe atoms are expected to exhibit favorable interactions with the BPN substrate.

Motivated by these considerations, in this work we employ first-principles calculations to systematically investigate the adsorption behavior of Fe atoms on 2×2 supercells of monolayer and bilayer BPN. Various adsorption configurations with different Fe coverages and adsorption sites are constructed, and low-energy adsorption configurations are identified through structural optimization of multiple initial structures. The corresponding average adsorption energies are subsequently evaluated. By combining adsorption-energy calculations with convex-hull analysis, we further examine the preferred adsorption configurations, coverage-dependent binding energetics, and relative thermodynamic stability of different Fe coverages against phase separation. On this basis, the mechanical properties of pristine and Fe-decorated BPN are investigated using the energy–strain method [15-18]. Particular attention is paid to the in-plane mechanical properties of Fe-adsorbed monolayer BPN and the out-of-plane elastic response of Fe-intercalated bilayer BPN. The in-plane elastic properties of pristine monolayer and bilayer BPN are also examined for comparison. Band structure calculations reveal that BPN exhibits typical metallic characteristics. Electronic transport is further examined within Boltzmann transport theory [19-21].

## 2. Computational methods

Density functional theory (DFT) calculations were carried out using the projector augmented wave method [22, 23] as implemented in the Vienna Ab initio Simulation Package[24-27] . The exchange-correlation energy was described by the Perdew–Burke–Ernzerhof functional [28] within the generalized gradient approximation. Long-range van der Waals interactions were treated using the DFT-D4 dispersion correction [29]. All the calculations were spin-polarized, with the plane-wave kinetic energy cutoff set at 500 eV. All the results reported in the main text were obtained without a Hubbard correction. As a robustness check, additional DFT+U calculations with $U$ = 2.0 eV applied to the Fe 3d states were performed for selected representative Fe adsorption configurations, as discussed in Sec. 3.1 and **Supplementary Sec. 1**. The self-consistent electronic iterations were considered converged when the total energy difference between successive steps was less than $10^{-5}$ eV.

Structural relaxations were performed until the Hellmann–Feynman forces on each atom were below $10^{-3}$ eV·Å$^{-1}$. To suppress spurious interactions between periodically repeated slabs along the out-of-plane direction, a vacuum layer of 14 Å was introduced along the direction perpendicular to the surface. In the structural relaxations, the Brillouin zone was sampled with a $k$-point spacing of 0.15 Å$^{-1}$.

For each Fe loading, the adsorption configurations were explored using a multi-start random search combined with a sequential-growth strategy. For monolayer BPN, multiple initial structures were generated by randomly placing Fe atoms at different positions above the BPN surface. For bilayer BPN, the initial configurations included Fe atoms adsorbed exclusively on the top surface, intercalated between the two BPN layers, or distributed across both regions. These configurations were subsequently fully relaxed and compared to identify low-energy structures for each Fe loading and adsorption environment. In parallel, starting from the lowest-energy $Fe_n$ structure obtained, additional $Fe_{n+1}$ candidates were generated by placing one additional Fe atom at different randomly selected positions. The relaxed structures from both search routes were compared, and the lowest-energy configuration obtained for each Fe loading was used for subsequent analysis. Depending on the Fe loading, approximately 5–10 initial configurations were examined for each $n$, with additional trials performed when different relaxed structures exhibited comparable energies. To evaluate the relative stability of different Fe coverages, convex-hull analyses were performed using the total adsorption energy per BPN supercell. The lower convex hull was constructed as a function of $n$ to assess the energetic stability of homogeneous Fe coverages against coexistence of regions with different Fe loadings.

The elastic constants were evaluated using the energy–strain method. Small deformations were applied to the equilibrium structures, and the corresponding strain energies were calculated and fitted within the harmonic elastic regime. For an orthorhombic two-dimensional system, the strain energy can be expressed as

$$\Delta E / A_0 = \frac{1}{2} C_{11} \varepsilon_x^2 + \frac{1}{2} C_{22} \varepsilon_y^2 + C_{12} \varepsilon_x \varepsilon_y + 2 C_{66} \varepsilon_{xy}^2$$

where $A_0$ is the equilibrium in-plane area. $C_{11}$ and $C_{22}$ characterize the longitudinal stiffness along the $x$- and $y$-directions, $C_{12}$ describes the coupling between the two normal strains, and $C_{66}$ represents the in-plane shear stiffness. For bilayer systems, $C_{33}$ characterizes the resistance to deformation along the out-of-plane $z$-direction, with the equilibrium average interlayer spacing adopted as the effective thickness (see **Supplementary Sec. 5** for details). The elastic constants were obtained by fitting the calculated strain energies as functions of the applied strains [15-18].

Within semiclassical Boltzmann transport theory, the conductivity tensor can be written as

$$\vec{\vec{\sigma}} = \frac{e^2}{4\pi^3}\int \tau(-\frac{\partial f}{\partial E})\vec{V}\vec{V}d^3\vec{k}$$

where $E$ is the electron energy, $\tau$ is the carrier relaxation time, $\vec{V} = \partial E / \hbar\partial\vec{k}$ is the electron velocity, and $f = 1 / (1 + \exp[(E - E_F) / k_B T])$ is the Fermi-Dirac distribution. In the conductivity calculations, the Brillouin zone was sampled with a $k$-point spacing of 0.03 Å$^{-1}$. The electronic transport coefficients were evaluated within the constant relaxation time approximation. Because the carrier relaxation time depends on the underlying scattering mechanisms and may vary with Fe coverage, no specific constant value of $\tau$ was assumed in the present work. Instead, the calculated results are reported as $\sigma / \tau$, which reflects the electronic-band-structure contribution to transport. Possible variations in the relaxation time arising from electron–phonon, impurity, or disorder scattering are therefore not explicitly included in the present calculations. For the two-dimensional BPN systems, the calculated $\sigma / \tau$ values were converted to volumetric units using the effective physical thickness of the material rather than the full supercell height including the vacuum region. This treatment avoids an artificial dependence of the reported transport coefficients on the chosen vacuum spacing.

The main calculation data of this work are compiled in **Supplementary Sec. 8**.

## 3. Results and discussion

### 3.1 *Adsorption of Fe atoms on monolayer BPN*

The biphenylene network (BPN) possesses a unique two-dimensional atomic structure (**Fig. 1(a)**), consisting of a planar $sp^2$-hybridized carbon net with tetragonal, hexagonal, and octagonal rings. Its rectangular primitive cell, belonging to the 2D *Pmm* space group, has lattice constants $a$ = 4.52 Å and $b$ = 3.77 Å. BPN is a metallic two-dimensional carbon allotrope, as indicated by its energy bands (**Fig. 1(b)**). The calculated electronic structure of pristine BPN is consistent with previous theoretical studies, which reported its metallic character and highly anisotropic electronic dispersion originating from the non-hexagonal 4–6–8 carbon network [30, 31].

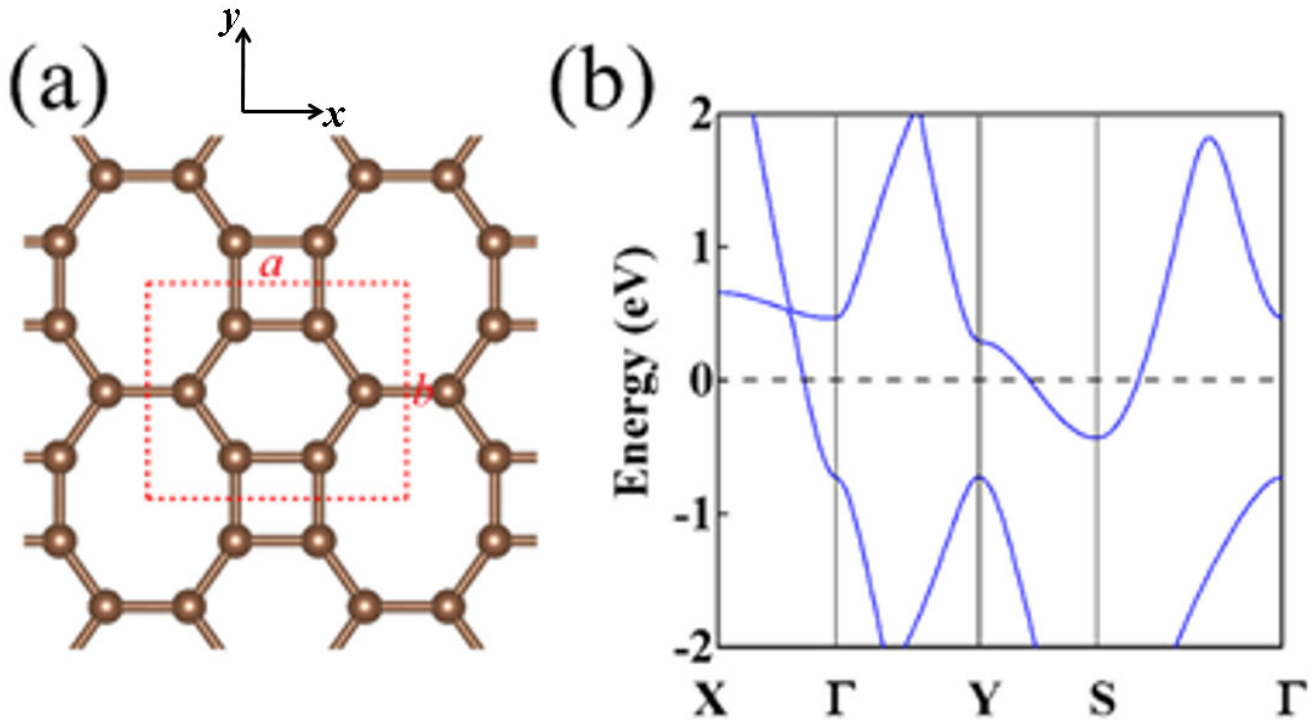

**Fig. 1** **(a)** The structure of monolayer BPN. The dashed lines represent the primitive cell. **(b)** The energy bands of monolayer BPN.

To investigate the adsorption energetics and relative stability of $Fe_n$ configurations on monolayer BPN, we first calculated the average adsorption energy per Fe atom, defined as

$$E_{ad}(n) = (\ E(\text{BPN-Fe}_n) - E(\text{BPN}) - nE(\text{Fe})\ ) \,/\, n$$

where $E(\text{BPN–Fe}_n)$ and $E$(BPN) are the total energies of the Fe-adsorbed and pristine BPN supercells, respectively, and $E$(Fe) is the energy of an isolated Fe atom. The average adsorption energy characterizes the binding strength per adsorbed Fe atom. To assess the thermodynamic stability of different Fe coverages against phase separation, the convex hull was constructed using the total adsorption energy per BPN supercell,

$$\Delta E(n) = nE_{ad}(n) = E(\text{BPN–Fe}_n) - E(\text{BPN}) - nE(\text{Fe})$$

The pristine substrate ($n$ = 0, $\Delta E$ = 0) was included as a reference state. Since the total adsorption energies are additive with respect to the substrate area, the lower convex hull of $\Delta E(n)$ versus $n$ provides an appropriate energetic criterion for coexistence of regions with different Fe coverages. Configurations lying on the hull are energetically favored against such phase separation within the set of sampled structures, whereas those above the hull are metastable with respect to mixtures of configurations on the hull. This analysis does not establish the absolute stability of Fe adsorption against all possible competing phases.

To investigate the adsorption of Fe atoms on BPN surface, we considered several configurations for $n$ Fe atoms on 2×2 BPN monolayer. Geometry relaxations were performed to identify the lowest-energy configurations obtained from our structural search. **Fig. 2(a)** presents the total adsorption energies $\Delta E(n)$ and the corresponding lower convex hull for Fe adsorption on 2×2 monolayer BPN. Within the investigated configurations, the convex hull connects pristine BPN ($n$ = 0) and the $Fe_{22}$-covered structure. All intermediate Fe coverages lie above this tie line, indicating an energetic preference for coexistence of Fe-poor and Fe-rich regions rather than homogeneous intermediate coverages, within the sampled adsorption configurations. This tendency is consistent with the increasingly favorable average Fe binding at higher loadings, although the average adsorption energies exhibit some nonmonotonic variations.

At low Fe coverage, the preferred adsorption site depends on the supercell size and the corresponding adsorbate concentration. To further understand it, the Fe–C bond lengths and Bader charges were analyzed. For the 2×2 BPN supercell, the single Fe atom preferentially occupies the center of the six-membered ring, with an Fe–C bond length of approximately 2.07 Å and a Bader electron population of 7.32 *e*. In contrast, for the larger 3×3 and 6×6 supercells, the preferred adsorption site shifts to the center of the four-membered ring, where the Fe–C bond length decreases slightly to approximately 2.02 Å and the Bader electron population increases to 7.41 *e*. The relatively small changes in both quantities indicate that the site transition is not associated with a substantial change in charge transfer or Fe–C bonding. Instead, it likely reflects the subtle energetic competition between different local coordination environments, which is affected by the Fe coverage and the periodic interaction between adsorbates. At sufficiently low coverage, the results converge toward the four-membered-ring site as the preferred adsorption configuration.

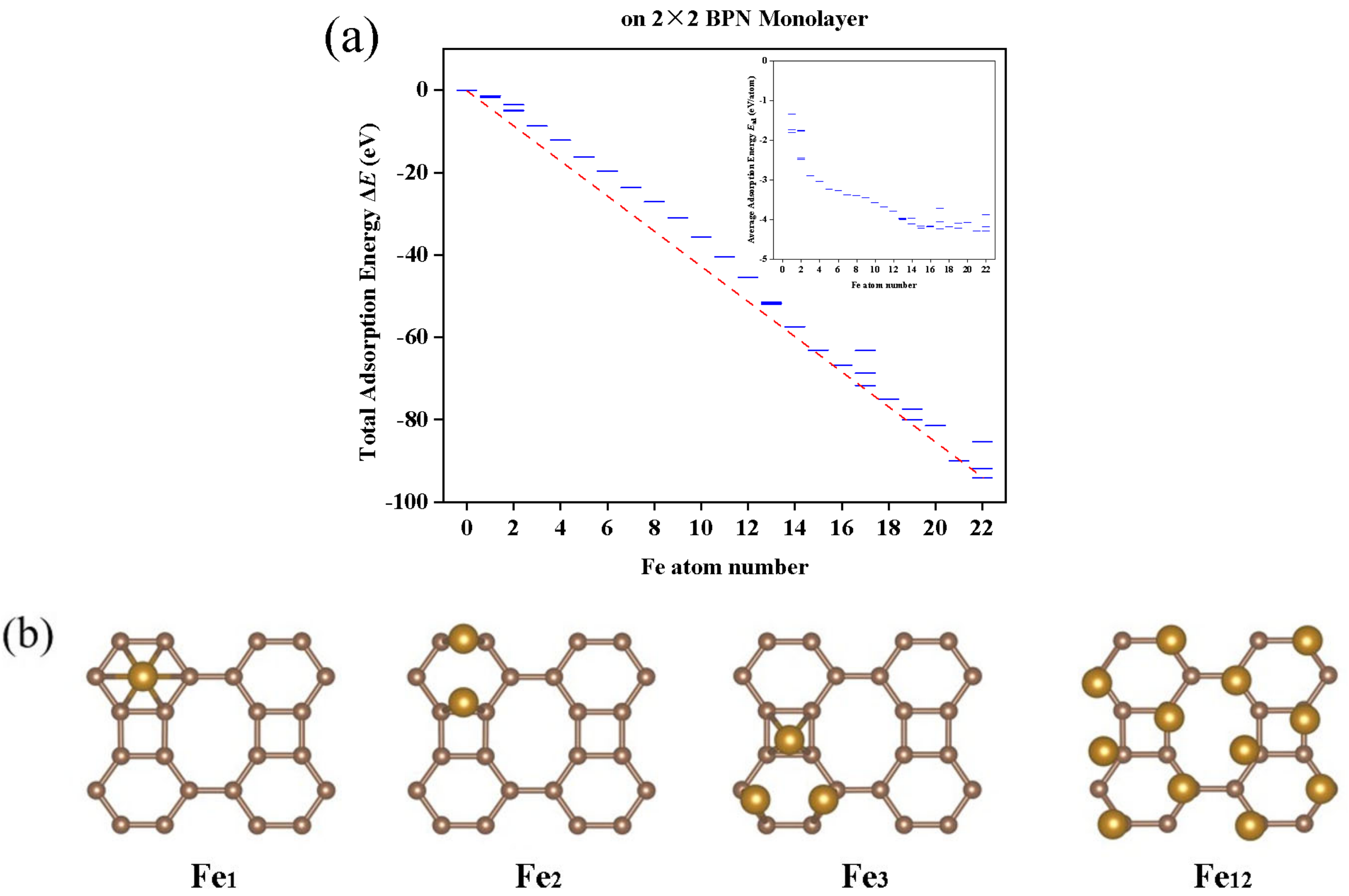

**Fig. 2 (a)** Total adsorption energy $\Delta E(n)$ of $Fe_n$ configurations on 2×2 monolayer BPN as a function of Fe loading. The dashed line represents the lower convex hull constructed by including pristine BPN ($n$ = 0). The inset shows the corresponding average adsorption energy per Fe atom, $E_{ad}(n)$. **(b)** Representative optimized Fe adsorption configurations on monolayer BPN.

To examine whether the treatment of Fe 3d localization affects the adsorption-energy trend, we additionally performed DFT+U calculations with $U = 2.0$ eV for the relevant adsorption configurations. The comparison shows that the inclusion of the Hubbard correction modifies the average adsorption energies of the selected configurations but does not substantially alter their relative energetic trends (**Supplementary Sec. 1**). This suggests that the qualitative adsorption-energy behavior of these representative configurations is not strongly affected by the Hubbard correction. Since Fe atoms contain partially filled 3d orbitals, the local magnetic moments and spin-resolved projected density of states (PDOS) were examined. The results are summarized in **Supplementary Sec. 2**, confirming the spin-polarized electronic characteristics of Fe/BPN systems.

When two Fe atoms are adsorbed, the average adsorption energy decreases significantly compared to the single-atom case, and the Fe atoms tend to aggregate at two positions on opposite edges of the six-membered ring, on the same side of the BPN sheet (**Fig. 2(b)**). As the Fe loading increases, the average adsorption energy generally becomes more negative, indicating increasingly favorable average Fe binding. At higher Fe coverages, the optimized structures exhibit increasingly pronounced Fe aggregation and structural rearrangements. As illustrated in **Supplementary Fig. S7**, the $Fe_{12}$ configuration already contains Fe atoms at different heights above the BPN surface, while further Fe incorporation promotes more extensive three-dimensional aggregation. Notably, the $Fe_{12}$ configuration lies above the corrected convex hull and is therefore energetically unfavorable against phase separation into Fe-poor and Fe-rich regions within the sampled configurations. For $n > 12$, the average adsorption energy exhibits a nonmonotonic dependence on the Fe loading rather than decreasing continuously. As shown in **Fig. 2(a)** and **Table S4**, $E_{ad}$ becomes less negative at $n = 16$, 18, and 20 relative to neighboring sampled coverages, followed by a more negative value at $n = 21$. These variations indicate that the average Fe binding strength depends sensitively on the specific atomic arrangement and structural relaxation at high coverages. The high-coverage structures suggest an increasing contribution from Fe–Fe interactions and the development of three-dimensional Fe aggregates on the BPN surface.

The incremental adsorption energy, defined as $E_{inc}(n) = \Delta E(n) - \Delta E(n-1)$, also exhibits variations with Fe loading. In particular, the relatively large negative value at $n = 13$ (approximately −6.6 eV) coincides with noticeable structural rearrangements between the $Fe_{12}$ and $Fe_{13}$ configurations (**Fig. S7**). This suggests that collective structural relaxation may contribute to the enhanced incremental binding at this coverage. Nevertheless, the $Fe_{13}$ configuration lies above the corrected convex hull and therefore does not represent a thermodynamically stable coverage against phase separation within the sampled states. Additional structural searches based on both forward Fe addition and reverse Fe removal did not identify lower-energy $Fe_{12}$ or $Fe_{13}$ configurations, lending further support to the reliability of the selected structures within the explored configuration space (**Supplementary Sec. 9**).

### 3.2 *Adsorption of Fe atoms on bilayer BPN*

To further understand the adsorption behavior of Fe atoms in a multilayer environment, we extended our analysis from BPN monolayer to bilayer. Three types of adsorption configurations were examined systematically: Fe atoms located exclusively in the interlayer region, solely on the top layer, and distributed across both the top layer and interlayer sites. As shown in **Fig. 3(a)**, the adsorption characteristics on BPN bilayer exhibit several similarities to those on BPN monolayer. For instance, when adsorption occurs only on the top layer, the most favorable adsorption sites closely resemble those identified in the monolayer case. Within the investigated Fe coverages ($n \leq 12$), no pronounced out-of-plane Fe aggregation is observed in the optimized bilayer configurations.

The convex-hull analysis (**Fig. 3(a)**) based on the total adsorption energies $\Delta E(n)$ identifies $n = 0$, 6, 8, and 12 as the hull vertices among the sampled bilayer configurations. All Fe-containing configurations on the hull correspond to interlayer adsorption, confirming that interlayer incorporation is energetically preferred over surface adsorption within the investigated configurations. The $Fe_6$ configuration also exhibits the most negative average adsorption energy, $E_{ad} = -4.50$ eV, among the sampled bilayer structures. The presence of $Fe_8$ and $Fe_{12}$ on the hull further indicates that these coverages are energetically competitive against phase separation into other sampled adsorption states.

The variation of average adsorption energy $E_{ad}(n)$ with the number of Fe atom reveals a clear hierarchy among different adsorption environments. For the same Fe loading, interlayer adsorption consistently exhibits the lowest (most negative) average adsorption energy, followed by the mixed "top-layer & interlayer" configuration, whereas adsorption solely on the top layer is the least energetically favorable. Notably, unlike the monolayer case, where the average adsorption energy decreases appreciably with increasing Fe coverage, Fe adsorption in the bilayer remains energetically favorable over the entire range of Fe concentrations considered. The dependence on Fe loading is non-monotonic, which reflects the competition among different adsorption configurations at discrete Fe loadings. Because the available adsorption sites in BPN are geometrically nonequivalent, increasing the number of Fe atoms changes the occupation of different sites as well as the Fe–Fe interactions and induces structural rearrangements. The particularly favorable configuration at $n = 6$ therefore results in a local minimum in the average adsorption energy, whereas further Fe addition does not necessarily lead to a monotonic enhancement of the average binding strength.

Moreover, our analysis shows that the four-membered ring centers within the interlayer region of the bilayer provide exceptionally stable adsorption sites. At low Fe coverages, the centers of the interlayer four-membered rings are among the most energetically favorable adsorption sites. Interestingly, pristine bilayer BPN exhibits an in-plane stacking offset of approximately 0.90 Å between the two layers. Upon interlayer Fe adsorption and structural relaxation, this offset is essentially eliminated, leading to a nearly aligned bilayer configuration (**Supplementary Sec. 3**).

**Fig. 3(b)** presents four representative configurations **a**, **b**, **c** and **d** with $n = 1, 6, 8$ and 12. The optimized interlayer distances are 3.51, 3.64, 3.67 and 3.98 Å for the $Fe_1$, $Fe_6$, $Fe_8$ and $Fe_{12}$ configurations, respectively. For a single Fe atom adsorbed in the interlayer of BPN bilayer, it preferentially occupies the center of a four-membered ring. When six Fe atoms are adsorbed, four occupy sites on both sides of the six-membered rings, followed by two at the bridge sites between adjacent six-membered rings, forming a nearly equilateral triangular arrangement. In the case of twelve Fe atoms, they are distributed relatively uniformly within the interlayer, with every three adjacent Fe atoms also forming an approximate equilateral triangle. In all four configurations, the intrinsic stacking offset of the BPN bilayer is effectively eliminated.

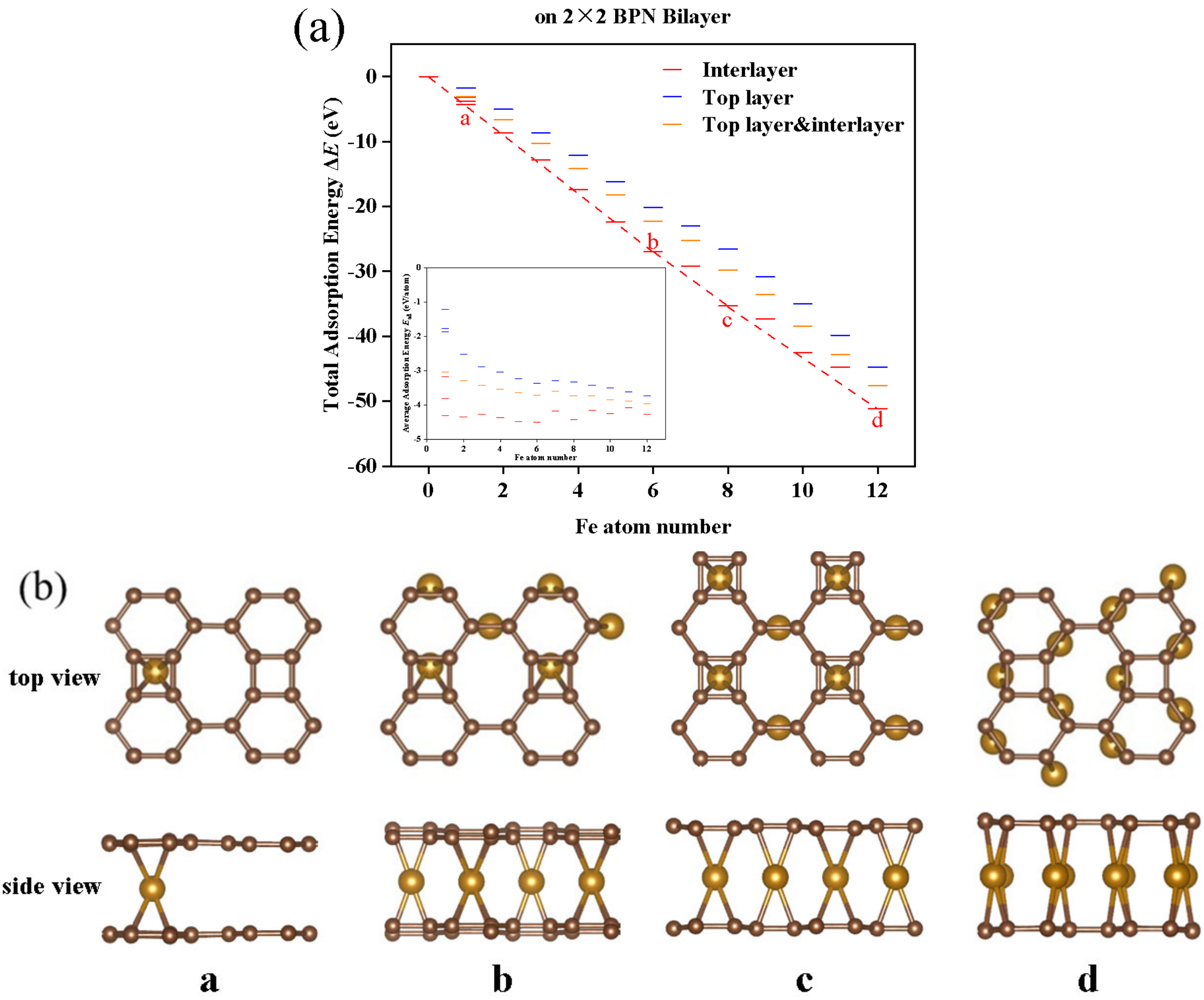

**Fig. 3 (a)** Total adsorption energies $\Delta E(n)$ of $Fe_n$ configurations on 2×2 bilayer BPN for top-layer, interlayer, and combined top-layer / interlayer adsorption. The dashed line represents the lower convex hull constructed from all sampled configurations, including pristine BPN ($n$ = 0). The inset shows the corresponding average adsorption energies per Fe atom, $E_{ad}(n)$. **(b)** Representative optimized Fe adsorption configurations on bilayer BPN.

For comparison with bulk Fe, we calculated the binding energy of α-Fe, which is −4.88 eV/atom. For $Fe_{22}$ on monolayer BPN, the average adsorption energy is −4.28 eV/Fe. This comparison indicates that Fe atoms are more strongly stabilized in bulk α-Fe than in the BPN-$Fe_{22}$ configuration. Together with the corrected convex-hull results, this finding suggests a thermodynamic driving force toward Fe aggregation and segregation at high loadings. However, the comparison does not establish the precipitation of bulk α-Fe on BPN, since finite-cluster surface energies, interfacial effects, and kinetic barriers are not explicitly accounted for.

3.3 *Effect of Fe adsorption on the mechanical properties of BPN*

The in-plane elastic properties of monolayer and bilayer BPN were systematically investigated using the energy–strain method. The calculated elastic constants for pristine monolayer BPN are $C_{11}$ = 243.86 N/m, $C_{12}$ = 84.50 N/m, $C_{22}$ = 303.67 N/m, and $C_{66}$ = 82.56 N/m, whereas those for pristine bilayer BPN are $C_{11}$ = 498.05 N/m, $C_{12}$ = 183.91 N/m, $C_{22}$ = 556.93 N/m, and $C_{66}$ = 167.13 N/m. All calculated elastic constants satisfy the Born–Huang mechanical stability criteria ($C_{11}C_{22} - C_{12}^2 > 0$, $C_{66} > 0$) [32], confirming the intrinsic mechanical stability of both monolayer and bilayer BPN. The calculated interfacial normal stiffness of bilayer BPN is 73.18 GPa/nm.

Based on these elastic constants, the orientation-dependent in-plane Young's modulus and Poisson's ratio were further evaluated by [33]

$$Y(\theta) = \left( \frac{C_{22}\cos^4\theta - 2C_{12}\cos^2\theta\sin^2\theta + C_{11}\sin^4\theta}{C_{11}C_{22} - C_{12}^2} + \frac{\cos^2\theta\sin^2\theta}{C_{66}} \right)^{-1}$$

and

$$v(\theta) = \frac{C_{12}C_{66}(\cos^4\theta + \sin^4\theta) + (C_{11}C_{22} - C_{12}^2 - C_{66}(C_{11} + C_{22}))\cos^2\theta\sin^2\theta}{C_{66}(C_{22}\cos^4\theta + C_{11}\sin^4\theta) + (C_{11}C_{22} - 2C_{12}C_{66} - C_{12}^2)\cos^2\theta\sin^2\theta}$$

For monolayer BPN, the two-dimensional Young's moduli along the crystallographic *x*- and *y*-directions are 220.3 N/m and 274.4 N/m, respectively. The Poisson's ratios $v_{xy}$ and $v_{yx}$ are 0.28 and 0.35, respectively. In bilayer BPN, the Young's moduli increase along both crystallographic directions, reaching values approximately twice and 1.8 times those of the monolayer along the *x*- and *y*-directions, respectively. The Poisson's ratios exhibit relatively modest changes, while the in-plane shear modulus is approximately doubled. These results indicate that bilayer formation enhances the in-plane stiffness while largely preserving the intrinsic mechanical anisotropy of BPN.

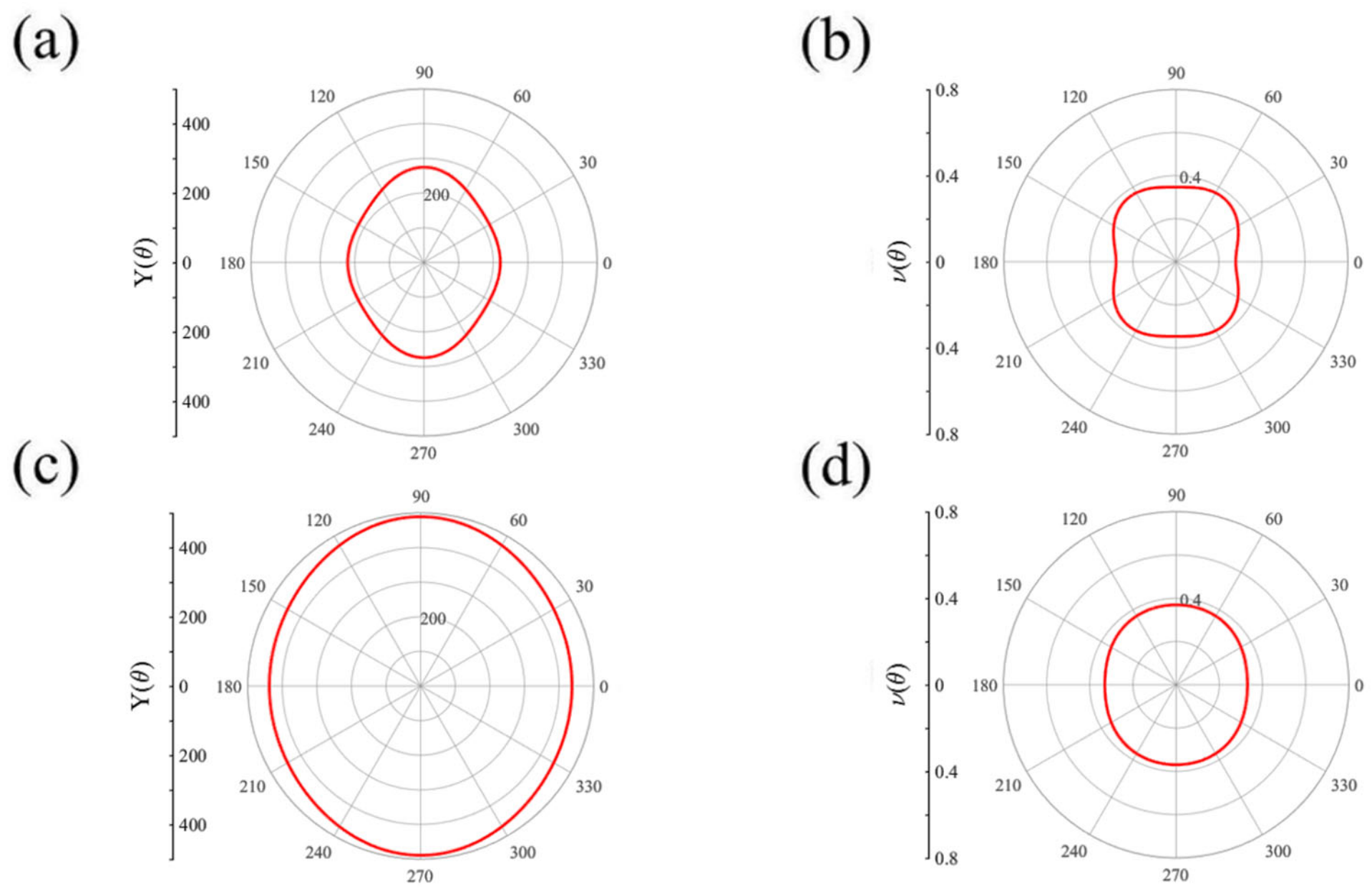

**Fig. 4** Polar plots of the orientation-dependent mechanical properties of 2×2 BPN supercells: **(a)** Young's modulus of monolayer BPN, **(b)** Poisson's ratio of monolayer BPN, **(c)** Young's modulus of bilayer BPN, and **(d)** Poisson's ratio of bilayer BPN.

As shown in **Fig. 4**, both monolayer and bilayer BPN exhibit pronounced in-plane mechanical anisotropy, with the Young's modulus along the *y*-direction being consistently higher than that along the *x*-direction. The mechanical anisotropy is mainly associated with its intrinsically anisotropic carbon network. As can be directly seen from the atomic structure in **Fig. 1(a)**, the 4–6–8 carbon network of BPN is not structurally equivalent along the x and y directions. The different orientations and connectivities of the C–C bonds therefore naturally lead to different in-plane elastic responses along the two crystallographic directions. Overall, the relatively high Young's modulus and shear modulus demonstrate that BPN possesses excellent resistance to both tensile deformation and shear distortion. The out-of-plane mechanical behavior of bilayer BPN was also examined. The calculated out-of-plane elastic constant $C_{33}$ is 24.59 GPa, which is notably lower than that of single-crystal graphite (36.50 GPa) [34]. This suggests that bilayer BPN is comparatively softer along the z-direction and more susceptible to out-of-plane deformation, reflecting the relatively weak interlayer interaction within the pristine bilayer structure.

Furthermore, the mechanical properties of Fe-adsorbed monolayer and bilayer BPN were systematically investigated. For monolayer BPN, the relevant mechanical response is mainly associated with deformation of the two-dimensional carbon network, and we therefore focus on its in-plane elastic properties. In bilayer BPN, the interlayer coupling is mainly governed by weak van der Waals interactions. Fe intercalation introduces much stronger interactions between the two BPN layers and therefore can markedly modify the out-of-plane mechanical response. Accordingly, our mechanical analysis focuses on the in-plane properties of monolayer BPN and, for bilayer BPN, particularly on the out-of-plane properties that are most sensitive to the Fe-induced modification of interlayer coupling. For monolayer BPN, as shown in **Fig. 5(a)**, the in-plane Young's moduli exhibit nonmonotonic variations with increasing Fe coverage, without a clear systematic enhancement in stiffness. In particular, at $n = 12$, $Y(0)$ and $Y(\pi/2)$ decrease to 200.1 and 229.0 N/m, respectively, corresponding to reductions of approximately 9.2% and 16.5% relative to pristine BPN. Nevertheless, the directional difference remains robust, with $Y(\pi/2)$ consistently higher than $Y(0)$ for all investigated Fe coverages. These results indicate that Fe adsorption modifies the in-plane stiffness while preserving the intrinsic mechanical anisotropy of the BPN framework.

The in-plane Poisson's ratios of Fe-adsorbed monolayer BPN were also evaluated from the calculated elastic constants, with $v_{xy} = C_{12} / C_{22}$ and $v_{yx} = C_{12} / C_{11}$. As shown in **Fig. S6**, the Poisson's ratios exhibit moderate fluctuations with Fe coverage, without a clear monotonic dependence on the number of Fe atoms. The directional difference in the Poisson's ratio further reflects the anisotropic in-plane elastic response of the BPN-based systems.

For bilayer BPN, using the equilibrium average interlayer spacing of each bilayer system as its effective thickness, the out-of-plane elastic constant $C_{33}$ was evaluated from the energy–strain relation (for calculation detail, see **Supplementary Sec. 5**).

The results are shown in **Fig. 5(b)**. The interlayer spacing of pristine bilayer BPN is 3.36 Å, yielding $C_{33}$ = 24.59 GPa. This value is lower than the out-of-plane elastic constant of single-crystal graphite ($C_{33}$ = 36.50 GPa) [34], indicating a relatively weak resistance to out-of-plane deformation. For comparison, the out-of-plane elastic constant of bulk BPN is calculated to be $C_{33}$ = 24.55 GPa, which is very close to that of bilayer BPN. This similarity suggests that bilayer and bulk BPN possess comparable out-of-plane mechanical characteristics.

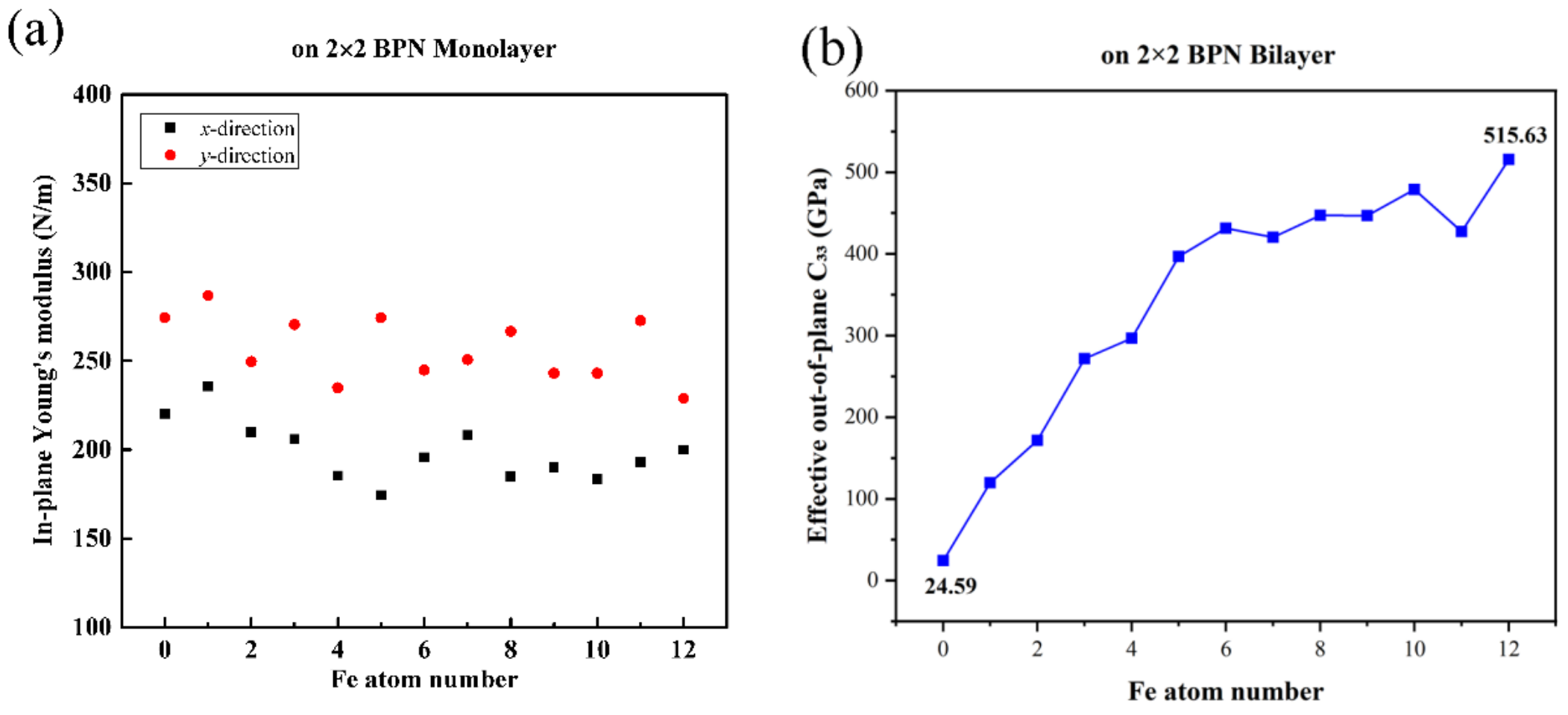

**Fig. 5 (a)** Variation of the in-plane Young's modulus of 2×2 monolayer BPN with Fe atom adsorption. **(b)** Variation of the effective out-of-plane elastic constant $C_{33}$ of Fe-intercalated 2×2 bilayer BPN.

When Fe atoms are introduced into the interlayer region of bilayer BPN, the out-of-plane elastic response can be substantially enhanced. For the representative $Fe_{12}$-intercalated 2×2 BPN bilayer, the average interlayer spacing increases from 3.36 Å to 3.98 Å, while the corresponding $C_{33}$ increases from 24.59 GPa to 515.63 GPa. This pronounced enhancement cannot be attributed to a reduced interlayer distance. Structural analysis shows that the interlayer Fe atoms are approximately fourfold coordinated with neighboring C atoms, with typical Fe–C distances of about 2.1 Å, and simultaneously interact with carbon atoms from both BPN layers. The charge-density-difference analysis further reveals evident charge accumulation in the Fe–C interfacial regions, while the Fe 3d and C 2p states exhibit appreciable overlap in the occupied states near the Fermi level (**Supplementary Sec. 6**). These results indicate that Fe intercalation introduces Fe-mediated interlayer bonding interactions, which bridge the two BPN sheets and provide additional resistance against relative out-of-plane displacement. The dramatic increase in $C_{33}$ therefore originates primarily from this Fe-mediated interlayer coupling rather than from geometrical contraction of the bilayer.

3.4 *Effect of Fe adsorption on the conductivity of BPN*

To further elucidate the influence of Fe adsorption on the electronic transport behavior of BPN, the electrical conductivity normalized by the relaxation time $\sigma / \tau$ was systematically evaluated for pristine and Fe-adsorbed monolayer/bilayer BPN systems. All transport calculations were performed at 300 K to enable a consistent comparison among different Fe coverages. The optimized lowest-energy configurations identified for monolayer BPN at each Fe loading were used for the corresponding transport calculations. For bilayer BPN, transport calculations were performed for the three adsorption environments considered in this work — top-layer adsorption, interlayer adsorption, and combined top-layer/interlayer adsorption—to compare their distinct effects on electronic transport.

The calculated results reveal that BPN exhibits pronounced anisotropic electrical transport behavior, with $\sigma / \tau$ along one crystallographic direction being significantly higher than that along the perpendicular direction. Such anisotropy originates from the intrinsic anisotropic bonding topology and electronic dispersion characteristics of BPN. For monolayer BPN, as shown in **Fig. 6(a)**, $\sigma / \tau$ exhibits a nonmonotonic evolution with increasing Fe adsorption concentration, initially decreasing and subsequently increasing at higher Fe loadings. Within the constant relaxation time approximation, the nonmonotonic variation of $\sigma / \tau$ originates from Fe-induced modifications of the electronic structure rather than changes in carrier scattering. To further understand this behavior, the density of states (DOS) near the Fermi level was examined for representative Fe coverages (**Supplementary Sec. 4**). The DOS profiles show pronounced variations near the Fermi level with increasing Fe coverage, indicating that Fe adsorption modifies the electronic states involved in carrier transport. These electronic-structure changes provide a possible origin for the coverage-dependent variation of $\sigma / \tau$.

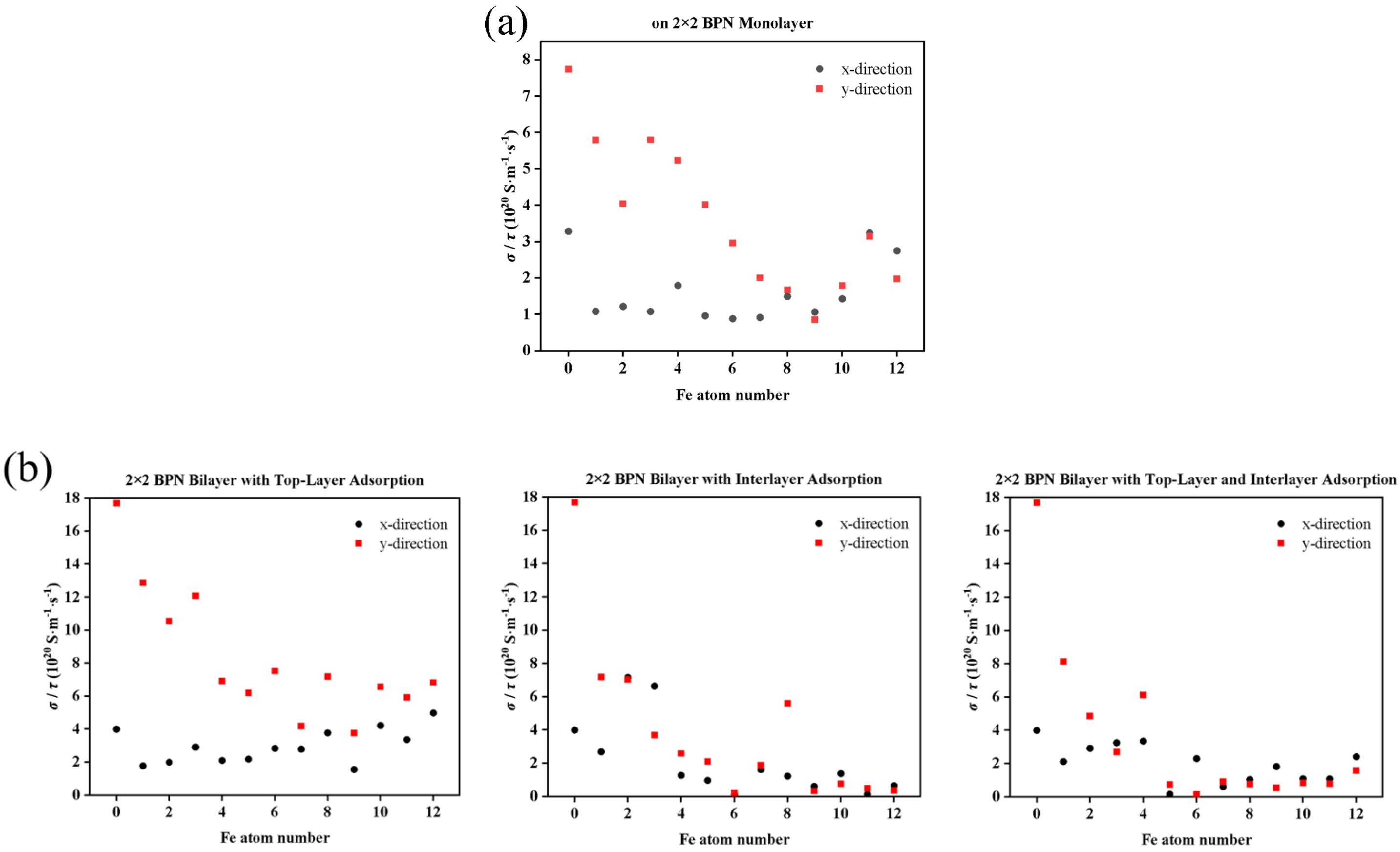

**Fig. 6** Electrical conductivity normalized by the relaxation time ($\sigma / \tau$) of **(a)** monolayer and **(b)** bilayer BPN with Fe adsorption.

The calculated $\sigma / \tau$ results for bilayer BPN are presented in **Fig. 6(b)**. In contrast, when Fe atoms are introduced into the interlayer region, either alone or together with surface adsorption, $\sigma / \tau$ is generally reduced over a broad range of Fe coverages compared with the top-layer adsorption case, although the detailed coverage dependence differs between the two configurations. This indicates that interlayer Fe adsorption exerts a more substantial influence on the electronic transport properties of bilayer BPN than surface adsorption. This difference can be associated with the distinct electronic-structure modification induced by interlayer Fe adsorption. Unlike surface adsorption, interlayer Fe atoms interact simultaneously with carbon atoms from both BPN layers, giving rise to Fe–C hybridized states and charge redistribution across the interlayer region. These changes can modify the electronic states contributing to transport and therefore lead to transport characteristics different from those of the top-layer adsorption case.

The calculated $\sigma / \tau$ values exhibit strong configuration-dependent variations. For the $Fe_6$-intercalated bilayer, $\sigma_{xx}/\tau$ and $\sigma_{yy}/\tau$ decrease by approximately 97.0% and 98.8%, respectively, relative to pristine bilayer BPN. These results demonstrate that interlayer Fe incorporation can strongly suppress the calculated transport coefficients, although the magnitude depends on the adsorption configuration and Fe coverage. The absolute conductivity would additionally depend on the actual carrier relaxation time and associated scattering mechanisms.

## 4. Summary

In summary, we systematically investigated the adsorption behavior of Fe atoms on monolayer and bilayer BPN and their effects on the mechanical and electrical properties using DFT calculations. The average adsorption energy on monolayer BPN generally becomes more negative with increasing Fe loading, although nonmonotonic variations are observed at higher coverages. At higher Fe coverages, the optimized structures exhibit increasingly pronounced three-dimensional Fe aggregation, accompanied by structural rearrangements of the adsorbed Fe atoms. The corrected convex-hull analysis, based on the total adsorption energies, indicates an energetic preference for coexistence of Fe-poor and Fe-rich regions rather than homogeneous intermediate coverages within the sampled monolayer configurations. For bilayer BPN, interlayer adsorption is energetically preferred over surface adsorption, with $n$ = 6, 8, and 12 identified as convex-hull vertices among the investigated Fe-containing configurations. These findings reveal distinct coverage-dependent adsorption energetics and structural evolution in monolayer and bilayer BPN.

The mechanical calculations reveal that both monolayer and bilayer BPN possess excellent in-plane mechanical stability, characterized by relatively high Young’s and shear moduli together with pronounced in-plane anisotropy. For monolayer BPN, Fe adsorption induces nonmonotonic variations in the in-plane Young's moduli, with no systematic stiffening observed as the Fe coverage increases. Although the in-plane stiffness is reduced at certain Fe coverages, the Young's modulus along the *y*-direction remains consistently higher than that along the *x*-direction, indicating that the intrinsic mechanical anisotropy of BPN is preserved after Fe adsorption. In contrast, the out-of-plane mechanical response of bilayer BPN is highly sensitive to interlayer Fe adsorption. The out-of-plane elastic constant $C_{33}$ increases dramatically from 24.59 GPa for pristine bilayer BPN to 515.63 GPa after Fe incorporation, demonstrating that interlayer Fe atoms can effectively strengthen interlayer coupling and substantially enhance the out-of-plane rigidity of the bilayer structure.

Transport calculations show that Fe adsorption substantially modifies both the magnitude and directional anisotropy of $\sigma / \tau$. The anisotropy varies nonmonotonically with Fe coverage, and the dominant transport direction reverses at certain coverages. The absolute conductivity is expected to additionally depend on the carrier relaxation time and the associated scattering mechanisms, which are not explicitly treated within the present constant-relaxation-time framework.

Overall, the present results demonstrate that Fe adsorption provides an effective approach for tuning the structural, mechanical, and electronic properties of biphenylene networks, particularly the interlayer mechanical behavior of bilayer BPN. These findings not only deepen the understanding of transition-metal interactions with BPN, but also highlight the considerable potential of Fe-decorated BPN for applications in nanoelectronics, low-dimensional conductive materials, and related functional devices.

**Funding Declaration**

This research did not receive funding.

**Conflict of Interest**

The authors declare that they have no conflicts of interest.

**Data Availability**

The data that support the findings of this study are available within the article and the supplementary.

**Supplementary Information**

# First-Principles Study of Fe Adsorption and Its Effects on the Mechanical and Electrical Properties of Monolayer and Bilayer Biphenylene Networks

Xiao-Ke Zhang, Zheng-Zhe Lin*

*School of Physics, Xidian University, Xi'an 710071, China*

* Corresponding Author. E-mail address: zzlin@xidian.edu.cn

## 1. Comparison of PBE and PBE+U

All primary results presented in this work were obtained using PBE without a Hubbard correction. Additional PBE+U calculations were performed for selected configurations to assess the sensitivity of the adsorption energetics to the treatment of Fe 3d electrons. We considered that the localization of 3d electrons in Fe systems requires careful consideration. Particularly, for strongly correlated systems such as transition-metal oxides or isolated low-coordination Fe species, the application of DFT+U or hybrid functionals is often necessary. However, in BPN-Fe systems, the Fe atoms are strongly coupled with the conductive BPN layer, and the Fe–Fe interactions lead to a metallic Fe-like environment rather than isolated Fe atoms. Therefore, DFT+U may over-localize the Fe 3d states and lead to an overestimation of magnetic moments.

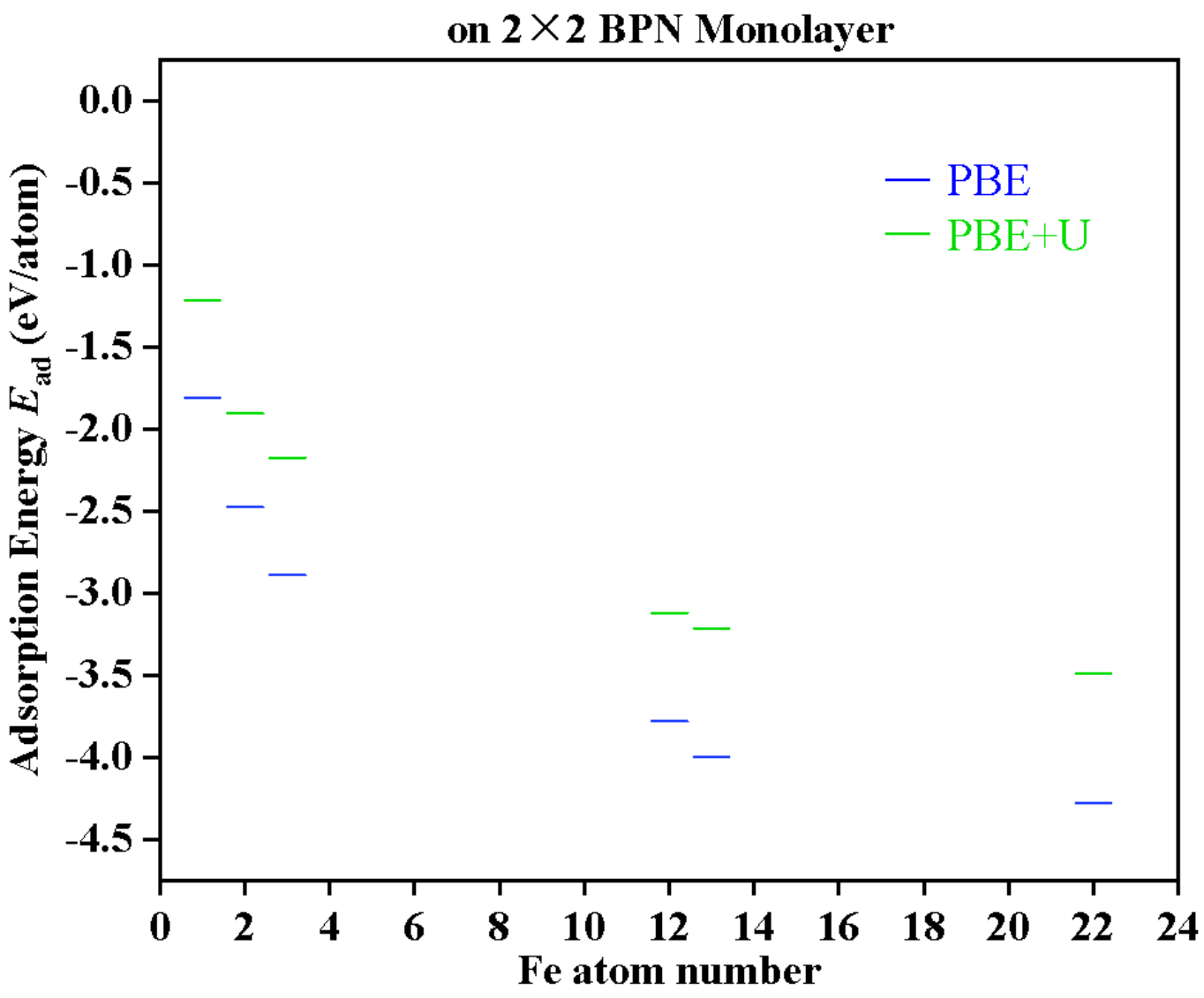

**Fig. S1** Comparison of PBE and PBE+U on the average adsorption energy of Fe atoms on 2×2 single-layer BPN.

To assess the sensitivity of the adsorption energetics to the treatment of Fe 3d electrons, we compared PBE and PBE+U ($U$ = 2.0 eV) results for selected Fe adsorption configurations on monolayer BPN. As shown in **Fig. S1**, the Hubbard correction generally shifts the average adsorption energies upward, while the relative energetic trends of the selected configurations remain similar. This comparison provides an assessment of the sensitivity of the adsorption energies to the Hubbard correction, rather than an independent verification of the convex-hull stability of all investigated Fe coverages.

In addition, we compared the derived atomic magnetic moment of α-Fe bulk (2.21 $\mu_B$ for PBE, 2.76 $\mu_B$ for PBE+U). The result of PBE is close to the experimental value (2.22 $\mu_B$), while PBE+U overestimates the magnetic moment. BPN-Fe systems with a large number of Fe atoms are conductive and close to Fe bulk. Therefore, PBE was adopted for the main calculations, while the additional PBE+U results were used to assess the sensitivity of selected adsorption energies.

## 2. Magnetic moments of Fe atoms

To investigate the magnetic properties, **Table S1** and **S2** show the magnetic moments of Fe atoms on 2×2 BPN monolayer and bilayer, respectively. For monolayer BPN, the average magnetic moment per Fe atom approaches the bulk α-Fe value at higher Fe loadings, although the dependence is nonmonotonic. On BPN bilayer, some Fe atoms are surrounded by more C atoms and thus the 3d electrons are less localized, which leads to smaller magnetic moments.

**Table S1** The average magnetic moment of Fe atoms on 2×2 BPN monolayer.

| Fe atom number | 1 | 2 | 3 | 12 | 13 | 14 | 22 |
|---|---|---|---|---|---|---|---|
| Magnetic moment ($\mu_B$) | 1.93 | 2.94 | 2.67 | 2.70 | 2.65 | 2.57 | 2.41 |

**Table S2** The average magnetic moment of Fe atoms on 2×2 BPN bilayer.

| Fe atom number | 1 | 2 | 3 | 6 | 12 | 14 | 18 |
|---|---|---|---|---|---|---|---|
| Magnetic moment ($\mu_B$) | 1.32 | 1.81 | 1.42 | 1.33 | 1.35 | 1.65 | 1.77 |

**Fig. S2** plots PDOS of $Fe_{12}$ on BPN monolayer and $Fe_6$ on BPN bilayer (the representative configurations). The results show that spin polarization mainly occurs in Fe 3d orbitals.

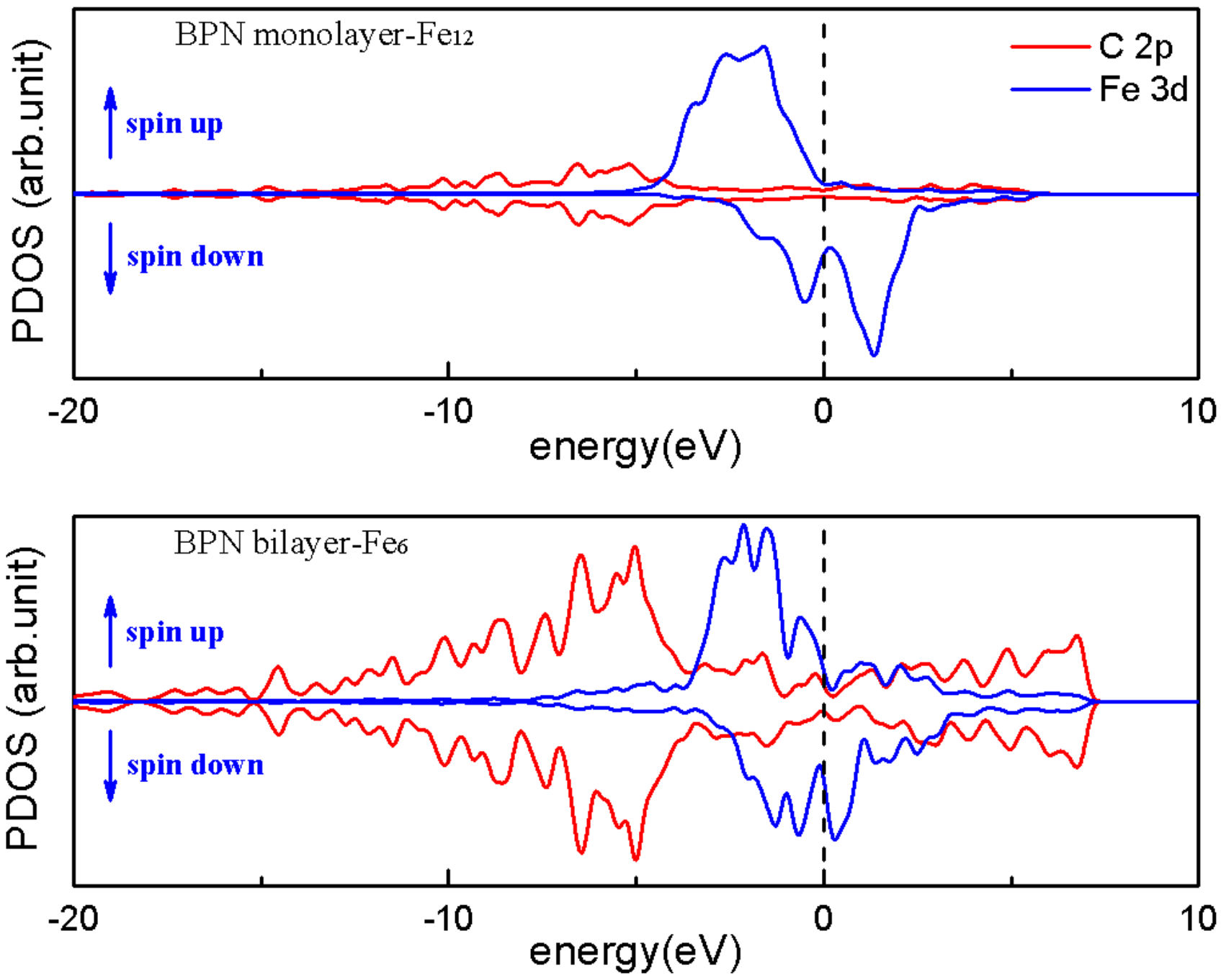

**Fig. S2** PDOS of $Fe_{12}$ on BPN monolayer (upper panel) and $Fe_6$ on BPN bilayer (lower panel). The Fermi level is set to zero.

### 3. Stacking configuration of BPN bilayer

**Fig. S3** shows pristine BPN bilayer and 2×2 BPN bilayer with a Fe atom. The optimized pristine BPN bilayer exhibits a laterally shifted stacking configuration. The two BPN layers are displaced relative to each other by approximately 0.90 Å along the in-plane direction, rather than adopting an exactly eclipsed stacking configuration. After the insertion of an Fe atom into the interlayer region, the two BPN layers undergo a relative lateral displacement during structural relaxation. As a result, the initial stacking offset is essentially eliminated, and the two layers become nearly aligned. This structural rearrangement indicates that the strong interaction of the interlayer Fe atom with both BPN layers modifies the interlayer registry and stabilizes the nearly aligned configuration.

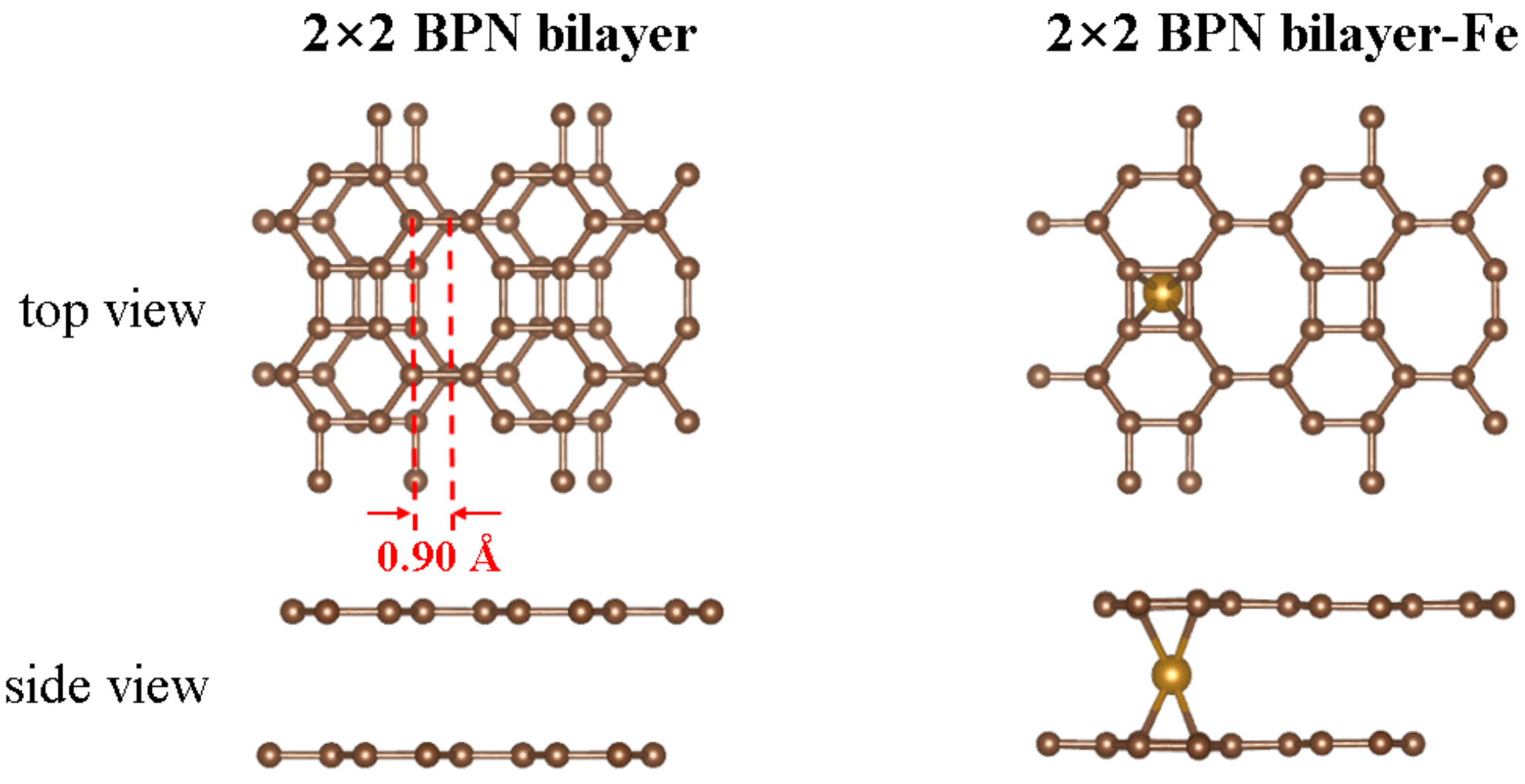

**Fig. S3** pristine BPN bilayer and 2×2 BPN bilayer with a Fe atom.

## 4. DOS of $Fe_n$ on BPN monolayer and bilayer

To illustrate the evolution of the electronic structure with increasing Fe coverage, DOS calculations were performed for representative low, intermediate, and high Fe coverages on BPN monolayer and bilayer. **Fig. S4(a)** shows the DOS of $Fe_1$, $Fe_{12}$, and $Fe_{22}$ adsorbed on a 2×2 BPN monolayer, while **Fig. S4(b)** presents the corresponding results for $Fe_1$, $Fe_6$, and $Fe_{12}$ on a 2×2 BPN bilayer. These results provide representative electronic-structure information across the investigated Fe-coverage range.

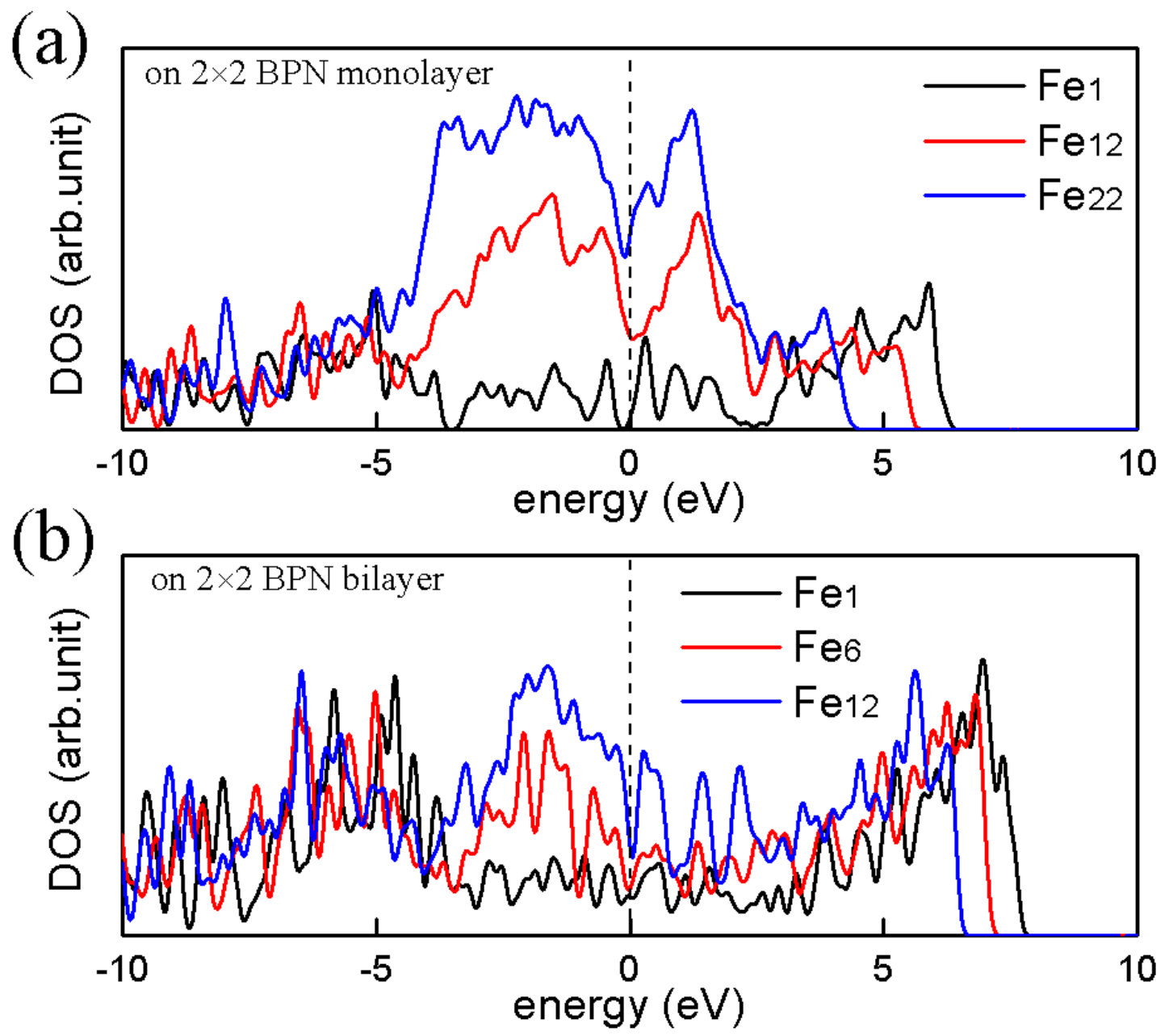

**Fig. S4** Total DOS of $Fe_n$ on **(a)** 2×2 BPN monolayer and **(b)** bilayer. The Fermi level is set to zero.

## 5. Calculation of the out-of-plane elastic constant $C_{33}$

The out-of-plane elastic constant $C_{33}$ was obtained using the energy–strain method. A series of uniaxial strains were applied along the *z*-direction, and the corresponding strain energies were fitted according to

$$\Delta E = \frac{1}{2} V_{\mathrm{eff}} C_{33}\, \varepsilon_z^2$$

where $\varepsilon_z$ is the applied out-of-plane strain and $V_{\mathrm{eff}}$ is the effective volume of the bilayer system. For two-dimensional layered materials, the effective volume was determined as

$$V_{\mathrm{eff}} = A_0 d_{\mathrm{eq}}$$

where $A_0$ is the equilibrium in-plane area and $d_{\mathrm{eq}}$ represents the equilibrium average interlayer spacing. The interlayer spacing was calculated from the difference between the averaged *z*-coordinates of the carbon atoms in the upper and lower BPN layers

$$d_{\mathrm{eq}} = |< z >_{\mathrm{upper}} - < z >_{\mathrm{lower}}|$$

The effective thickness was determined separately for each bilayer configuration. Therefore, the pristine bilayer BPN and Fe-intercalated bilayer BPN systems were evaluated using their respective equilibrium interlayer spacings rather than a common thickness parameter. The calculated interlayer spacings and corresponding $C_{33}$ values are summarized in **Table S3**.

**Table S3** Interlayer spacing and out-of-plane elastic constant $C_{33}$ of bilayer BPN systems.

| System | $d_{eq}$ (Å) | $C_{33}$ (GPa) |
|---|---|---|
| BPN bilayer | 3.36 | 24.59 |
| Fe12-intercalated 2×2 BPN bilayer | 3.98 | 515.63 |

## 6. Bonding in $Fe_{12}$-intercalated BPN bilayer

To clarify the origin of the strongly enhanced out-of-plane stiffness after Fe intercalation, the local Fe–C coordination and electronic interaction were further analyzed. In the $Fe_{12}$-intercalated 2×2 BPN bilayer, the interlayer Fe atoms are approximately fourfold coordinated with neighboring C atoms, with typical Fe–C distances of about 2.1 Å. Importantly, the Fe atoms interact simultaneously with carbon atoms from both the upper and lower BPN sheets, thereby forming Fe-mediated interlayer bridges.

The charge-density-difference map shown in **Fig. S5(a)** exhibits pronounced charge accumulation in the regions between Fe atoms and the neighboring C atoms, accompanied by charge depletion around the constituent atoms. This spatial charge redistribution indicates substantial Fe–C bonding interactions and supports the interpretation that the intercalated Fe atoms electronically couple the two BPN layers. The PDOS in **Fig. S5(b)** further shows an appreciable overlap between the Fe 3d and C 2p states over the occupied states near the Fermi level, consistent with Fe–C orbital interaction. These results suggest that Fe intercalation changes the interlayer interaction from predominantly weak van der Waals coupling in pristine bilayer BPN to Fe-mediated interlayer bonding, thereby providing additional resistance against relative out-of-plane displacement of the two BPN sheets and contributing to the pronounced enhancement of $C_{33}$.

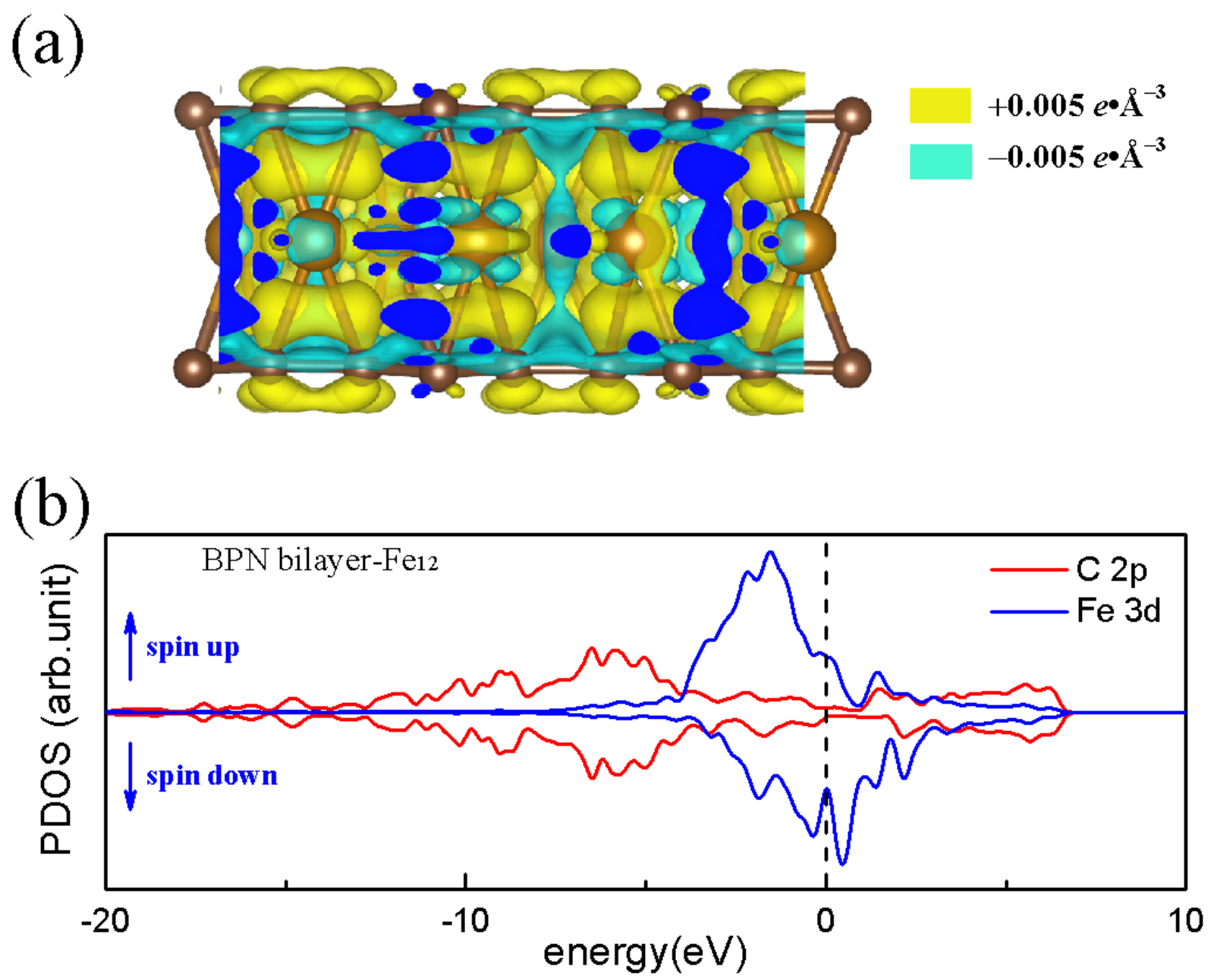

**Fig. S5 (a)** Charge density difference and **(b)** PDOS of $Fe_{12}$-intercalated 2×2 BPN bilayer.

## 7. Poisson's ratio of Fe-adsorbed monolayer BPN

The in-plane Poisson's ratios of Fe-adsorbed monolayer BPN were evaluated from the calculated elastic constants according to

$$v_{xy} = \frac{C_{12}}{C_{22}}, \qquad v_{yx} = \frac{C_{12}}{C_{11}}$$

where $v_{xy}$ denotes the transverse strain response along the $y$ direction under uniaxial loading along $x$, while $v_{yx}$ corresponds to the transverse response along $x$ under loading along $y$. As shown in **Fig. S6**, both Poisson's ratios vary moderately with the number of adsorbed Fe atoms, and no clear monotonic dependence on Fe coverage is observed. Nevertheless, the two directional Poisson's ratios remain generally different over the investigated coverage range, consistent with the anisotropic in-plane elastic response of the BPN framework. The Poisson's ratios remain positive for all considered Fe coverages.

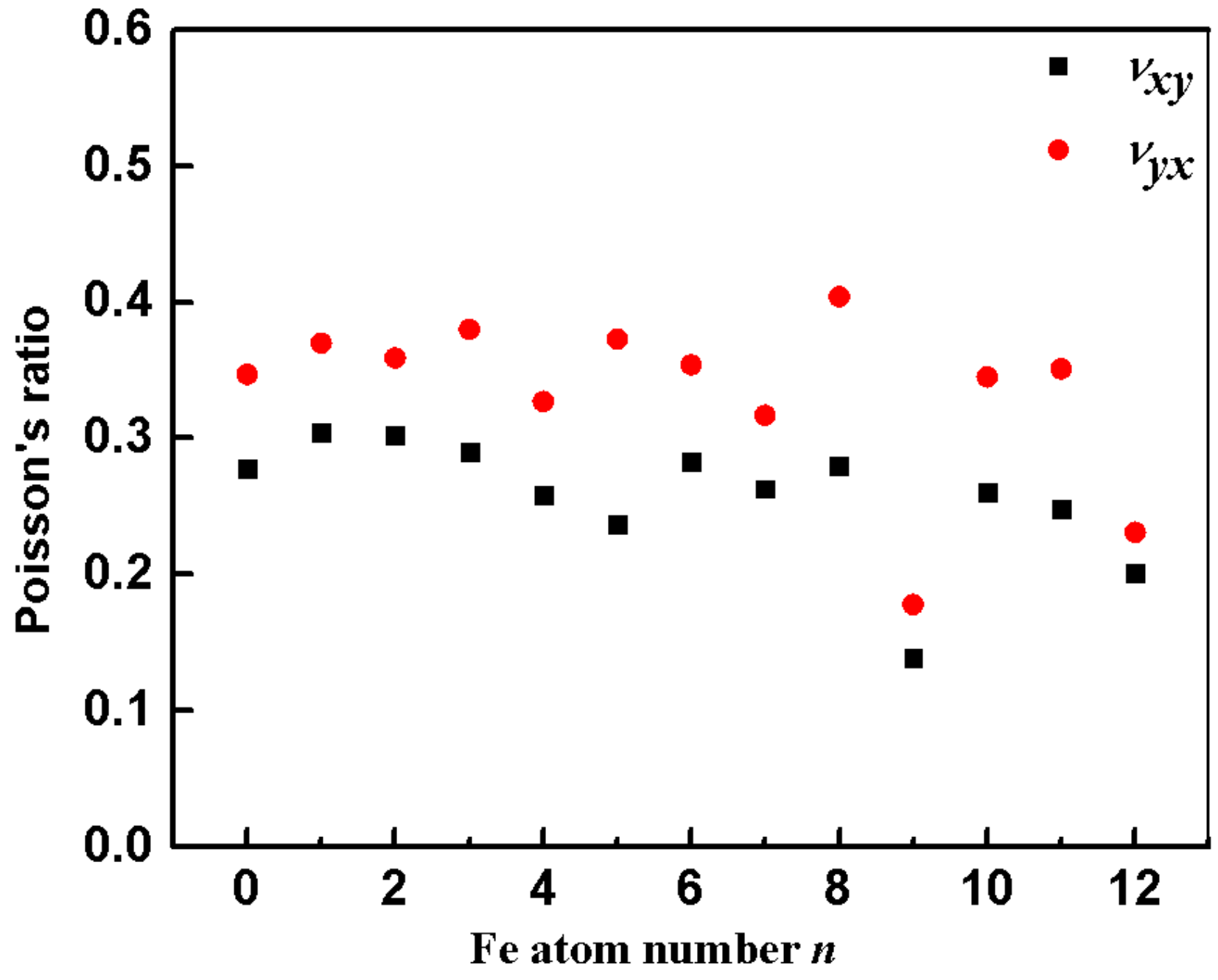

**Fig. S6** In-plane Poisson's ratio $\nu_{xy}$ and $\nu_{yx}$ of $Fe_n$-adsorbed 2×2 BPN monolayer.

## 8. Calculation data

The data of this paper are collected in the following tables. **Table S4** corresponds to **Fig. 2(a)**. **Table S5** corresponds to the Fe-intercalated configurations in **Fig. 3(a)**. **Table S6** corresponds to **Fig. 5(a)**. **Table S7** corresponds to **Fig. 5(b)**. **Table S8** corresponds to **Fig. 6(a)**. **Table S9**, **S10** and **S11** correspond to the three panels in **Fig. 6(b)** from left to right.

**Table S4** Average and total adsorption energies of $Fe_n$ on 2×2 BPN monolayer.

| Fe atom number $n$ | 1 | 2 | 3 | 4 | 5 | 6 | 7 | 8 | 9 | 10 | 11 |
|---|---|---|---|---|---|---|---|---|---|---|---|
| $E_{ad}$ (eV) | -1.81 | -2.48 | -2.89 | -3.02 | -3.24 | -3.27 | -3.38 | -3.39 | -3.45 | -3.56 | -3.68 |
| $\Delta E$ (eV) | -1.81 | -4.95 | -8.68 | -12.10 | -16.18 | -19.64 | -23.63 | -27.10 | -31.01 | -35.64 | -40.46 |

| Fe atom number $n$ | 12 | 13 | 14 | 15 | 16 | 17 | 18 | 19 | 20 | 21 | 22 |
|---|---|---|---|---|---|---|---|---|---|---|---|
| $E_{ad}$ (eV) | -3.78 | -4.00 | -4.11 | -4.21 | -4.17 | -4.22 | -4.17 | -4.22 | -4.07 | -4.27 | -4.28 |
| $\Delta E$ (eV) | -45.40 | -52.00 | -57.47 | -63.11 | -66.78 | -71.73 | -75.00 | -80.09 | -81.48 | -89.76 | -94.12 |

**Table S5** Average and total adsorption energies of Fe intercalated in 2×2 BPN bilayer.

| Fe atom number $n$ | 1 | 2 | 3 | 4 | 5 | 6 | 7 | 8 | 9 | 10 | 11 | 12 |
|---|---|---|---|---|---|---|---|---|---|---|---|---|
| $E_{ad}$ (eV) | -4.30 | -4.35 | -4.27 | -4.36 | -4.47 | -4.50 | -4.17 | -4.42 | -4.14 | -4.25 | -4.07 | -4.26 |
| $\Delta E$ (eV) | -4.30 | -8.69 | -12.81 | -17.43 | -22.35 | -26.98 | -29.19 | -35.33 | -37.29 | -42.47 | -44.76 | -51.17 |

**Table S6** In-plane Young's modulus $Y(0)$ along the $x$-direction and $Y(\pi/2)$ along the $y$-direction of 2×2 monolayer BPN with Fe atom adsorption.

| Fe atom number $n$ | 0 | 1 | 2 | 3 | 4 | 5 | 6 | 7 | 8 | 9 | 10 | 11 | 12 |
|---|---|---|---|---|---|---|---|---|---|---|---|---|---|
| $Y(0)$ (N/m) | 220.3 | 235.8 | 209.7 | 205.9 | 185.5 | 174.5 | 195.7 | 208.2 | 184.9 | 190.1 | 183.4 | 193.2 | 200.1 |
| $Y(\pi/2)$ (N/m) | 274.4 | 286.8 | 249.6 | 270.5 | 235.0 | 274.3 | 244.9 | 250.7 | 266.7 | 243.1 | 243.1 | 272.7 | 229.0 |

**Table S7** Out-of-plane elastic constant $C_{33}$ of Fe-intercalated 2×2 bilayer BPN.

| Fe atom number $n$ | 0 | 1 | 2 | 3 | 4 | 5 | 6 | 7 | 8 | 9 | 10 | 11 | 12 |
|---|---|---|---|---|---|---|---|---|---|---|---|---|---|
| $C_{33}$ (GPa) | 24.6 | 115.5 | 171.7 | 271.7 | 296.8 | 396.8 | 431.3 | 420.3 | 447.3 | 446.7 | 479.0 | 427.3 | 515.6 |

**Table S8** Electrical conductivity normalized by the relaxation time ($\sigma / \tau$) of monolayer BPN with Fe adsorption.

| Fe atom number $n$ | 0 | 1 | 2 | 3 | 4 | 5 | 6 | 7 | 8 | 9 | 10 | 11 | 12 |
|---|---|---|---|---|---|---|---|---|---|---|---|---|---|
| $\sigma_{xx} / \tau$ ($10^{20}$ S·m$^{-1}$·s$^{-1}$) | 3.28 | 1.08 | 1.22 | 1.08 | 1.79 | 0.96 | 0.88 | 0.91 | 1.49 | 1.07 | 1.43 | 3.24 | 2.75 |
| $\sigma_{yy} / \tau$ ($10^{20}$ S·m$^{-1}$·s$^{-1}$) | 7.75 | 5.80 | 4.05 | 5.80 | 5.23 | 4.02 | 2.96 | 2.01 | 1.68 | 0.85 | 1.79 | 3.14 | 1.98 |

**Table S9** Electrical conductivity normalized by the relaxation time ($\sigma / \tau$) of bilayer BPN with top-layer Fe adsorption.

| Fe atom number $n$ | 0 | 1 | 2 | 3 | 4 | 5 | 6 | 7 | 8 | 9 | 10 | 11 | 12 |
|---|---|---|---|---|---|---|---|---|---|---|---|---|---|
| $\sigma_{xx} / \tau$ ($10^{20}$ S·m$^{-1}$·s$^{-1}$) | 4.00 | 1.78 | 1.99 | 2.91 | 2.11 | 2.19 | 2.83 | 2.79 | 3.78 | 1.56 | 4.22 | 3.37 | 4.98 |
| $\sigma_{yy} / \tau$ ($10^{20}$ S·m$^{-1}$·s$^{-1}$) | 17.69 | 12.87 | 10.53 | 12.07 | 6.92 | 6.19 | 7.52 | 4.18 | 7.18 | 3.75 | 6.58 | 5.91 | 6.83 |

**Table S10** Electrical conductivity normalized by the relaxation time ($\sigma / \tau$) of bilayer BPN with interlayer Fe adsorption.

| Fe atom number *n* | 0 | 1 | 2 | 3 | 4 | 5 | 6 | 7 | 8 | 9 | 10 | 11 | 12 |
|---|---|---|---|---|---|---|---|---|---|---|---|---|---|
| $\sigma_{xx}$ / $\tau$ ($10^{20}$ S·m$^{-1}$·s$^{-1}$) | 4.00 | 2.69 | 7.16 | 6.65 | 1.27 | 0.97 | 0.12 | 1.62 | 1.23 | 0.60 | 1.38 | 0.14 | 0.65 |
| $\sigma_{yy}$ / $\tau$ ($10^{20}$ S·m$^{-1}$·s$^{-1}$) | 17.69 | 7.20 | 7.04 | 3.69 | 2.59 | 2.10 | 0.22 | 1.87 | 5.59 | 0.33 | 0.76 | 0.49 | 0.37 |

**Table S11** Electrical conductivity normalized by the relaxation time ($\sigma$ / $\tau$) of bilayer BPN with top-layer and interlayer Fe adsorption.

| Fe atom number *n* | 0 | 1 | 2 | 3 | 4 | 5 | 6 | 7 | 8 | 9 | 10 | 11 | 12 |
|---|---|---|---|---|---|---|---|---|---|---|---|---|---|
| $\sigma_{xx}$ / $\tau$ ($10^{20}$ S·m$^{-1}$·s$^{-1}$) | 4.00 | 2.12 | 2.92 | 3.25 | 3.35 | 0.16 | 2.30 | 0.61 | 1.04 | 1.82 | 1.09 | 1.08 | 2.41 |
| $\sigma_{yy}$ / $\tau$ ($10^{20}$ S·m$^{-1}$·s$^{-1}$) | 17.69 | 8.12 | 4.86 | 2.70 | 6.12 | 0.73 | 0.14 | 0.90 | 0.75 | 0.54 | 0.84 | 0.78 | 1.59 |

## 9. Structural evolution of Fe-adsorbed monolayer BPN at high coverages

**Fig. S7** presents top and side views of representative optimized Fe adsorption configurations on monolayer BPN at high coverages (*n* = 12, 13, 15, and 22). In these configurations, Fe atoms are located on one side of the BPN sheet. At *n* = 12, the Fe atoms already occupy different heights above the carbon framework. Further Fe incorporation leads to structural rearrangements and increasingly pronounced three-dimensional aggregation rather than uniform layer-by-layer growth. The $Fe_{13}$ configuration exhibits noticeable structural changes relative to $Fe_{12}$, while extensive Fe aggregation and deformation of the underlying BPN framework are observed at *n* = 22. These results highlight the complex interplay between Fe–C and Fe–Fe interactions at high Fe coverages.

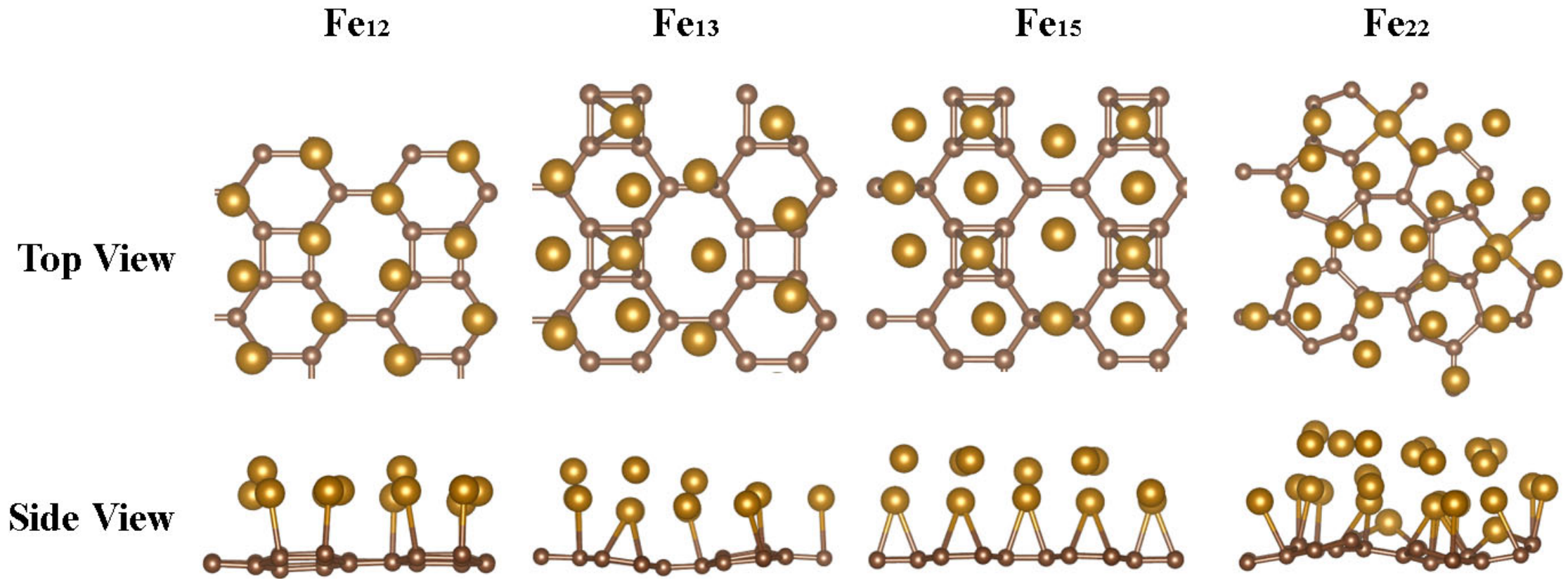

**Fig. S7** Top and side views of representative optimized Fe adsorption configurations on one side of monolayer BPN at high Fe coverages (*n* = 12, 13, 15, and 22).

To further assess the reliability of the optimized structures of *n* = 12 and 13, additional structural searches were performed. Several $Fe_{12}$ initial configurations were generated by removing Fe atoms from different positions in the optimized $Fe_{13}$ structure, while additional $Fe_{13}$ candidates were constructed by placing an Fe atom at different positions around the optimized $Fe_{12}$ structure. All candidates were subsequently relaxed using the same computational settings as in the original search. As summarized in **Table S12**, these additional searches did not identify configurations with lower energies than those previously obtained. This provides further support for the reliability of the selected $Fe_{12}$ and $Fe_{13}$ configurations within the explored structural space, although a global-minimum guarantee cannot be established.

**Table S12** Relative energies of $Fe_{12}$ and $Fe_{13}$ configurations obtained from the original and additional structural searches.

| System | Search route | Number of trials | Lowest relative energy (eV) |
|---|---|---|---|
| $Fe_{12}$ | Original search | 8 | 0.00 |
| $Fe_{12}$ | Removing one Fe from $Fe_{13}$ | 5 | 0.13 |
| $Fe_{13}$ | Original search | 8 | 0.00 |
| $Fe_{13}$ | Adding Fe to $Fe_{12}$ | 5 | 0.19 |